\documentstyle[epsfig]{aipproc}

\begin{document}
\title{Microquasars: hard X-ray/$\gamma$-ray emission
}
\author{Rob Fender}
\address{
Astronomical Institute `Anton Pannekoek' and Center for
High-Energy Astrophysics,  University of Amsterdam, Kruislaan 403,
1098 SJ Amsterdam, The Netherlands
}

\maketitle

\begin{abstract}

I review some of the basic observational details of jets from X-ray
binaries, or `microquasars'.  It is shown that in both (Z and Atoll)
NS and BHC systems radio emission, and therefore jet formation, is
correlated with the presence of hard (30--500 keV) X-ray emission. At
$\gamma$-ray ($>$ 500 keV) energies, the relation is not so clear.
Possible physical connections between the presence of a jet, with a
population of relativistic electrons, and the emission of hard X-rays,
are briefly discussed.

\end{abstract}

\section*{Introduction -- jets from X-ray binaries}  

The study of `microquasars', or the phenomenon of relativistic jets
from X-ray binary systems, is one of the most vigorously pursued
fields in observational high-energy astrophysics in recent years.
Once Mirabel \& Rodriguez (1994) had shown that X-ray binary systems
were capable of producing relativistic (bulk Lorentz factor $\Gamma
\geq 2$) outflows, whose primary signature was in the radio band, it
was recognised that such `jets' may constitute an important aspect of
the `accretion' process onto compact objects. In recent years
observations have clearly demonstrated the presence of jets in most
classes of X-ray binary, and furthermore have clearly indicated a
strong coupling with the accretion `state' of the source, whether
neutron star (e.g. Penninx et al. 1988) or black hole accretor
(e.g. Fender et al. 1999b). See e.g. Hjellming \& Han (1995), Mirabel
\& Rodriguez (1999), and other papers by this author for reviews of
radio emission and jets from X-ray binaries (although note that new
observations are coming thick and fast all the time!). See e.g. Lewin,
van Paradijs \& van den Heuvel (1995) for an excellent volume
containing reviews of the models for the different `states' of
accreting neutron stars and black holes (admittedly from the era
before the ubiquity and significance of jets was widely recognised).

\begin{figure}
\centering\epsfig{figure=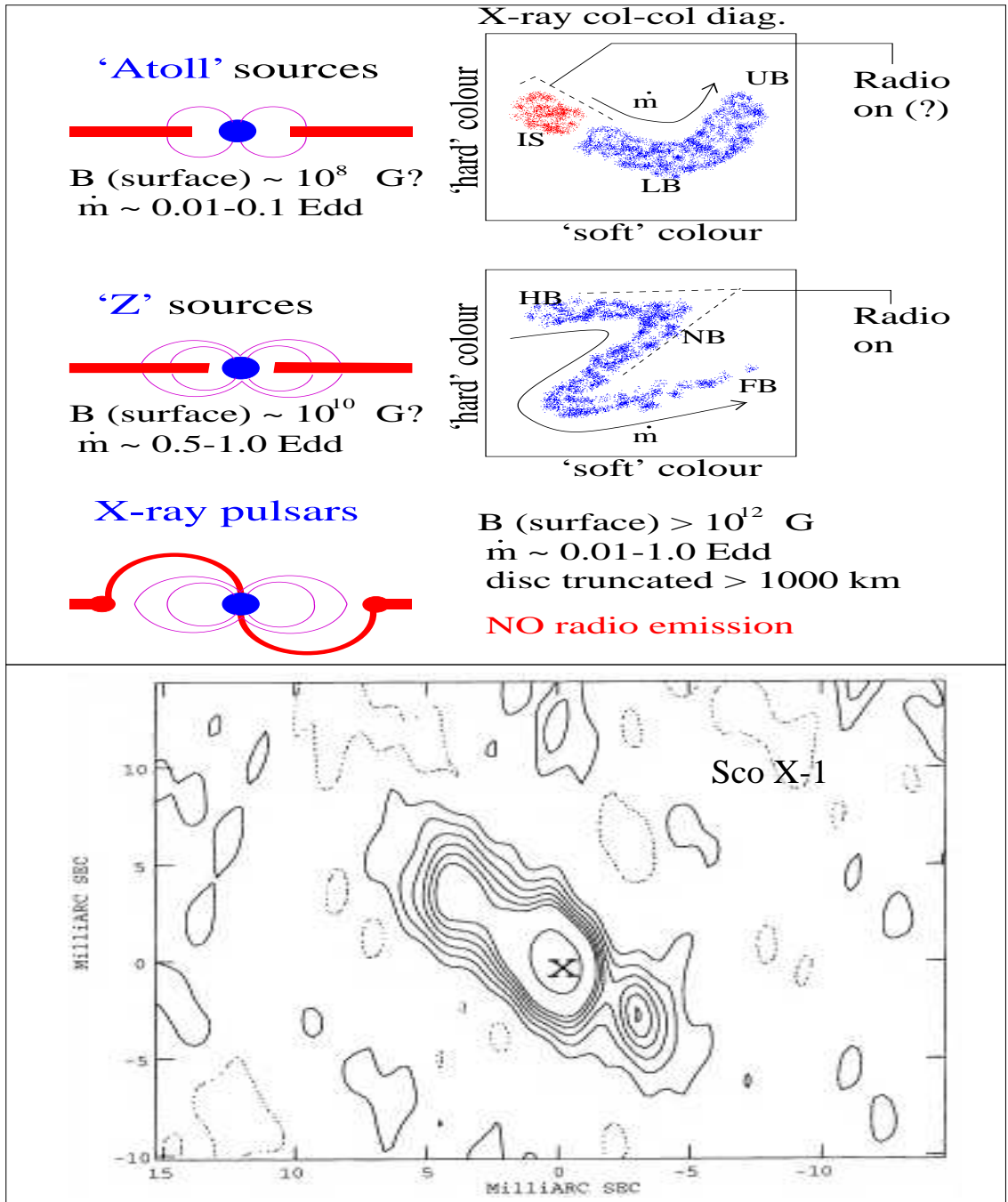,width=15cm,height=18cm,angle=0}
\caption{Top panel: a qualititative sketch of the relation of jets to
accretion in the three `types' of neutron star XRB. In the low-field
`Atoll' sources the accretion rate is believed to be $<10$\%
Eddington, except possibly during rare transient outbursts (e.g. Aql
X-1); the evidence is marginal so far but it appears such sources are
`radio on' when in the `Island State' (IS) in the X-ray colour-colour
diagram (CD). The Z sources are believed to be accreting at a much
higher rate, near Eddington, and are `radio on' when on the
`Horizontal Branch' and maybe also, at a lower level, the `Normal
Branch' in the CD. Note that for both Atoll and Z sources the
estimates of surface magnetic field are very uncertain. Finally, in
the high-field X-ray pulsars no radio synchrotron emission has ever
been detected; possibly this is due to truncation of the accretion
disc a long way from the neutron star. Lower panel: the jets of
the Z source Sco X-1, directly resolved with the VLBA (Bradshaw,
Fomalont \& Geldzahler 1999).}
\end{figure}

In Fig 1 and 2 I summarise the current
understanding of the empirical relation of radio emission to X-ray
`state' of the neutron star (NS) and black hole candidate (BHC)
systems respectively. In addition to the relations indicated in these
figures, it seems that discrete, radio-emitting ejection events (in
which the radio spectrum rapidly becomes optically thin) are
associated with state transitions (including outbursts of transients
-- Fig 3).

\begin{figure}
\centering\epsfig{figure=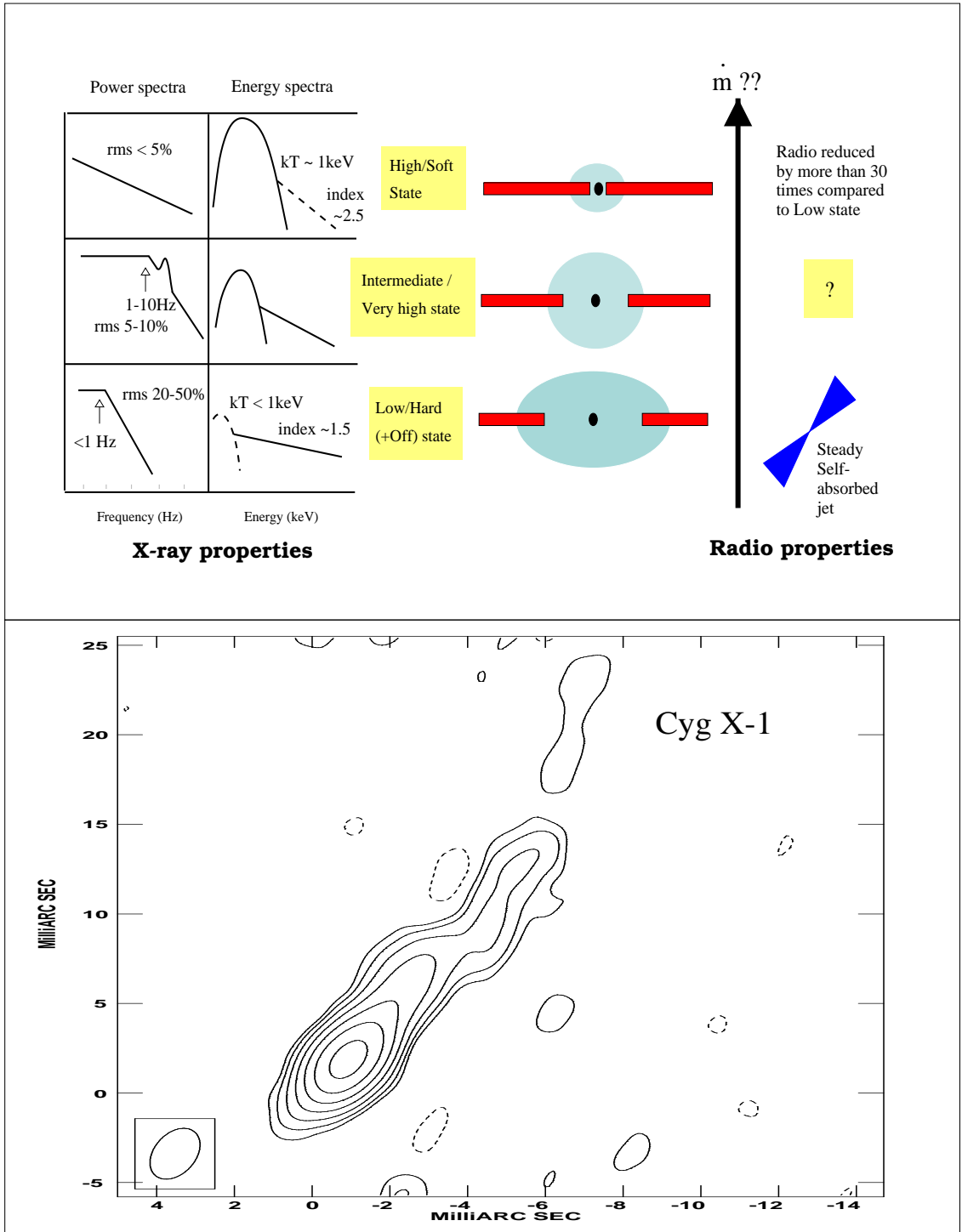,width=15cm,height=20cm,angle=0}
\caption{Top panel: the qualitative relation of radio emission to X-ray
`state' in BHC XRBs: the low/hard state is found to produce a steady,
flat- (or inverted-)spectrum jet, the soft state produces no
detectable radio emission. The relation of mass accretion rate, 
$\dot{m}$, to these states
is not certain (e.g. Homan et al. 2001); nor are the radio
characteristics 
of the relatively rare `Intermediate/Very High' states well determined.
The image in the lower panel is of the `steady' jet from the BHC Cyg
X-1, as resolved with the VLBA (Stirling et al. 2001). 
}
\end{figure}

While some areas of the coupling between accretion flow and jet are
still empirically uncertain (the Atoll sources, the Intermediate/Very
High state for BHCs), it seems that all XRBs except the high-field
X-ray pulsars will, under the right conditions, produce a
synchrotron-emitting jet.  It is therefore an important question to
address the significance of this jet, energetically and dynamically,
for the process of accretion onto compact objects as a whole. One
thing obvious from inspection of Figs 1(a),(b) is the apparent
anti-correlation between the mass accretion rate, $\dot{m}$,
traditionally estimated from X-ray studies alone, and the presence of
a jet, in both NS and BHC systems (see also Belloni, Migliari \&
Fender 2000 and Homan et al. 2001 for further discussion). The
energetics of jets from X-ray binaries in general, and BHCs in the
Low/Hard state in particular, are discussed in detail in Fender
(2001a,b), in which it is concluded that the jets probably contain at
least 20\% of the liberated accretion energy, maybe much more.

\section*{Hard X-rays : $\sim$30--500 keV}

In this energy band (rather arbitrarily defined here) there are
several detections of X-ray binary systems.

\subsection*{Neutron stars}

\begin{figure}
\centering\epsfig{figure=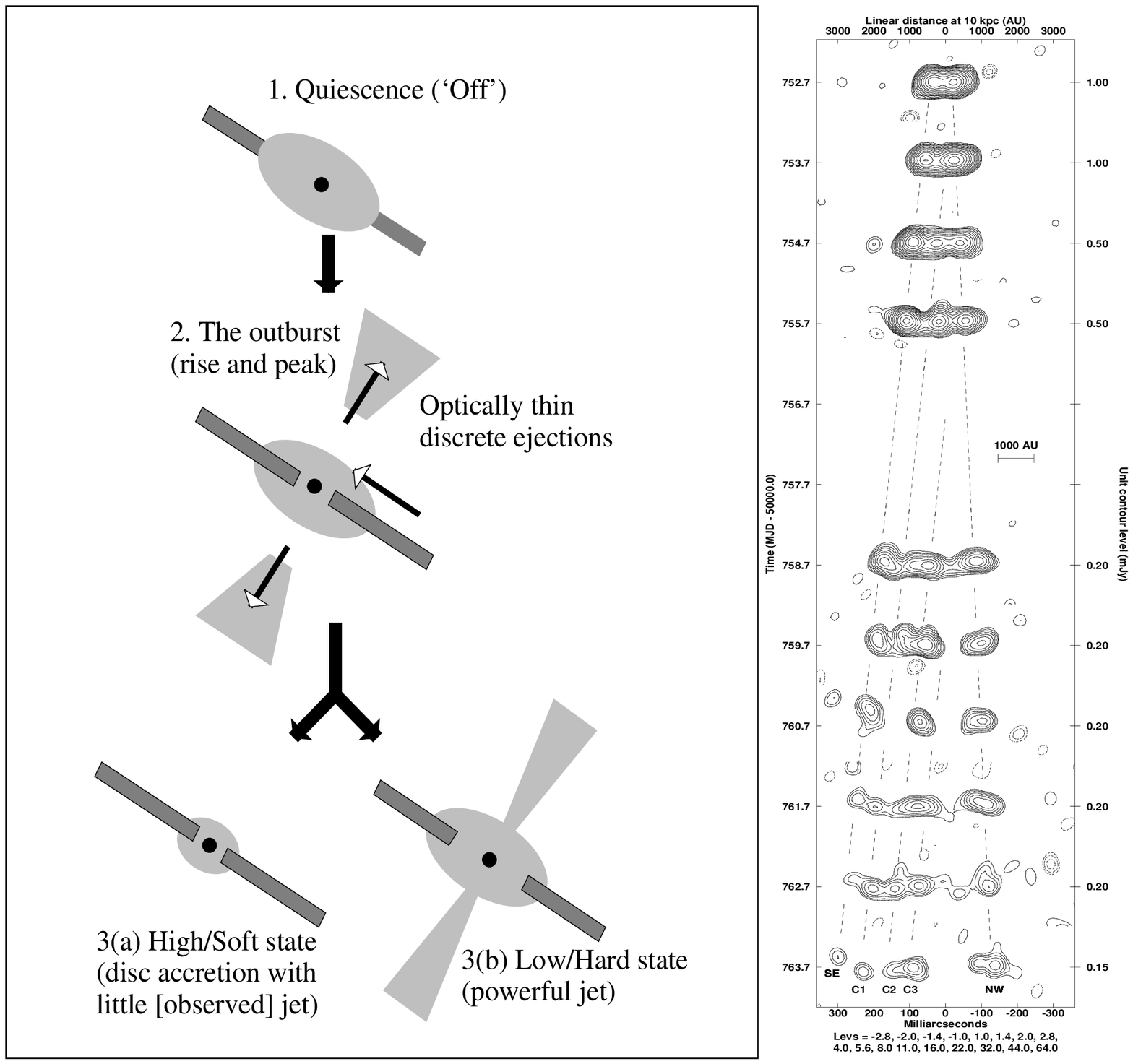,width=15cm,angle=0}
\caption{Left panel: optically thin radio events associated with
outbursts seem to arise in discrete ejections of material during the
rapid state change. In the case of black hole systems if the source
transits to the High/Soft X-ray state [stage 3(a)] the radio emission
remains optically thin and fades away; if the system instead transits
to the Low/Hard X-ray state [stage 3(b)] then a flat spectral
component, probably the signature of a powerful partially
self-absorbed jet, emerges (see Fig 2). Something like the sequence 
$1 \rightarrow 2 \rightarrow 3(a)$ may also occur in the NS
(atoll-like) transients.
Right panel: Resolved discrete relativistic
ejections from the BHC GRS 1915+105 with MERLIN (Fender et al. 1999a).}
\end{figure}

Both of the two major groups of neutron star systems which probably
produce jets, Z and
Atoll sources, have been detected in some cases to energies $\geq 30$
keV. In the case of the Z sources, a picture now seems to be emerging
whereby a hard tail, extending to $\geq 100$ keV, is present in
varying degree, being strongest on the Horizontal Branch (HB) of the
X-ray colour diagram (CD) and weakest (or non-existent) on the Flaring
Branch (FB; see Fig 1 for a schematic of a `typical' CD). Asai et
al. (1994) first found evidence for a hard tail from GX 5-1 which
decreased from the Normal Branch (NB) to the FB. More recently Di
Salvo et al. (2001 a,b) have reported the detection of a hard X-ray
tail from the Z sources GX 17+2 and GX 349+2; again the data seem to
be consistent with the hard X-ray tail being strongest in the HB/NB
and weakest on the FB. Iaria et al. (2001) also find evidence for a
hard tail in the jet source Cir X-1, which may be an unusual Z source.
Hard X-ray emission has also been reported from the archetypal Z
source Sco X-1, although its correspondence with position on the Z is
unclear (D'Amico et al. 2001). Strickman \& Barret (2000) report that
the hard X-ray emission in Sco X-1 may be correlated with periods of
radio flaring. Comparison of this picture with Fig 1 shows that
observationally, the presence of the hard tail seems to be correlated
with the presence of radio emission (and therefore almost certainly
jet production).  Note that in the received wisdom $\dot{m}$ increases
from HB$\rightarrow$NB$\rightarrow$FB, and so both jet and hard X-ray
production are anticorrelated with $\dot{m}$.

\begin{figure}
\centering\epsfig{figure=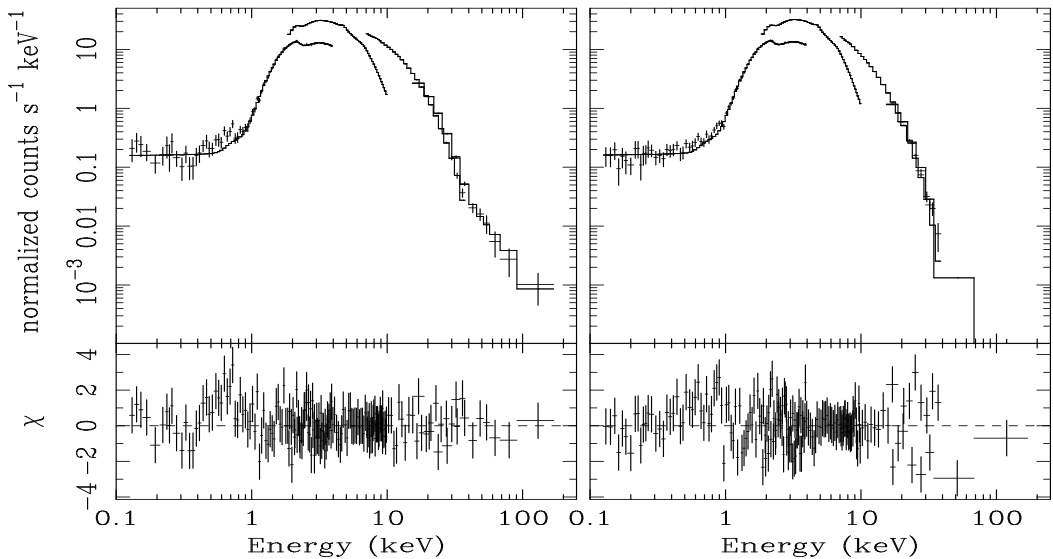,width=8cm,height=15cm,angle=270}
\caption{A hard X-ray tail from the Z source GX 17+2 (di Salvo et
al. 2000b); the left panel corresponds to the upper Horizontal Branch (HB),
the right panel to the lower Normal Branch (NB; see Fig 1 for a sketch
of the `Z' indicating HB, NB and `Flaring Branch' FB).
It is clear that the hard
($>30$ keV) X-ray excess is present only in the left spectrum; note
further that the Horizontal Branch the state corresponding to most
radio emission from Z sources (see Fig 1).
}
\end{figure}

The Atoll sources, weaker as a population than the Z sources at radio
wavelengths (Fender \& Hendry 2000), also display hard X-ray states
with emission detected up to and beyond 100 keV. For example, di Salvo et
al. (2000) report the detection to such high energies 
of the atoll source 4U 1728-34, and
Barret et al. (2000) report hard X-ray tails from four other Atoll-type
X-ray binaries. It is clear that in `low' states of the Atoll sources
a hard X-ray power-law is present which seems to be the dominant
component in the broadband spectral energy distribution. If, as we
suspect, both by analogy with BHCs and recent observations,
(work in preparation), that radio emission in Atoll sources is also
associated with `low' X-ray states, specifically the Island State (IS)
in the Atoll CD, we again find, as with the Z sources, that radio and
hard X-ray emission are correlated with each other and anti-correlated
with apparent $\dot{m}$.

\begin{figure}
\centering\epsfig{figure=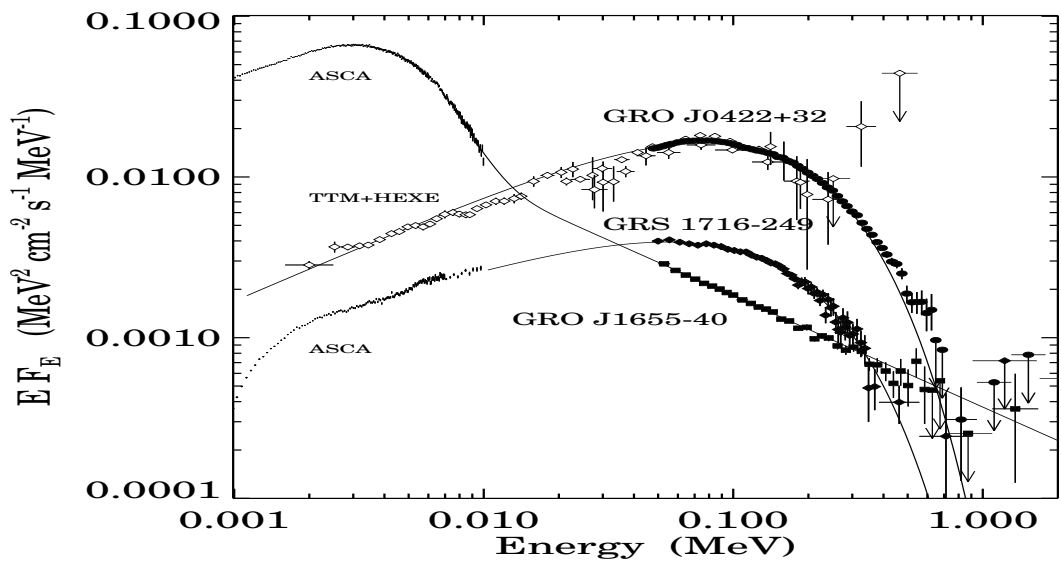,width=15cm,height=8cm,angle=0}
\caption{X-ray spectra of the BHCs GRO J0422+32 and GRS 1716-249 in
the Low/Hard state (which corresponds to `steady' jet production --
see e.g. Fig 2), with spectral energy distributions which peak around
100--200 keV. Also, the X-ray spectrum of 
the relativistic jet source GRO J1655-40, with a strong soft component
and a power-law tail which may extend to $\geq 1$ MeV.
From Grove et al. (1998).
}
\end{figure}

\subsection*{Black Holes}

Many BHC systems have been detected at energies above 30 keV; in fact
the X-ray power-law which dominates the high-energy radiation from
BHCs in the Low/Hard spectral state generally extends to at least 100
keV. Furthermore, although the High/Soft state is dominated
energetically by a `soft' disc component, there is increasing evidence
that a steep power-law extends beyond this thermal component to at
least several hundred keV (Grove et al. 1997, 1998; Poutanen 1998). Fig 2
illustrates the X-ray spectral and timing properties of the main
`states' schematically, and Fig 5 shows some examples of `real' X-ray
spectra. Hard X-ray emission seems to be a general, although not
exclusive, property of BHCs.

\begin{figure}
\centering\epsfig{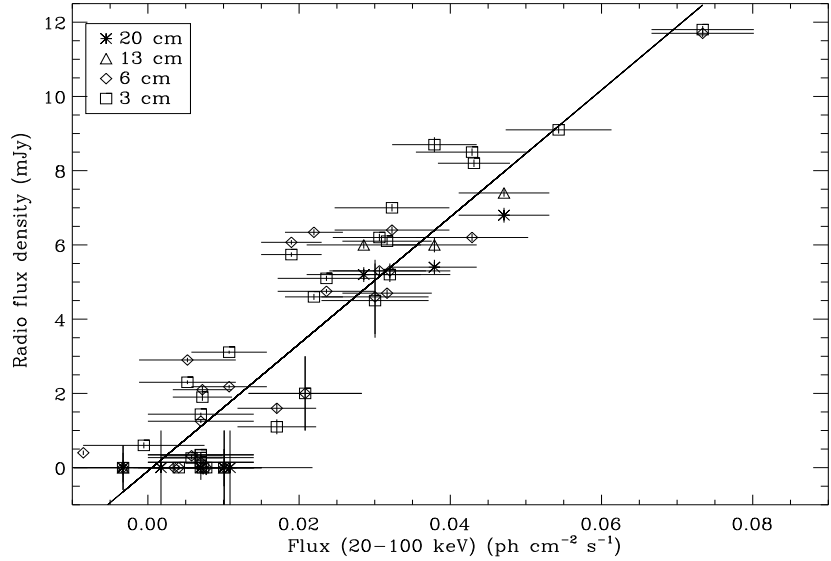}
\caption{Correlation between (GHz) radio and hard X-ray (20-100 keV,
as measured with BATSE) emission from the BHC GX 339-4. From Corbel et
al. (2000).
}
\end{figure}

The BATSE and OSSE instruments onboard the (late) CGRO mission contributed an
exceptional amount to our understanding of the relation of the hard
X-ray emission from BHCs (e.g. Grove et al. 1997, 1998). Brocksopp et
al. (1999) and Corbel et al. (2000) showed that there is a strong
correlation between hard X-rays (as measued with BATSE) and radio
emission in the persistent BHCs Cyg X-1 and GX 339-4 (Fig 6).
McCollough et al. (1999) showed that the jet source Cyg X-3 (which may
be a BHC) shows both correlated and anti-correlated radio -- hard
X-ray behaviour (Fig 7); this may correspond to different X-ray
`states' and/or rapid state transitions in this source (whose
accretion properties remain shrouded by a dense wind).  The BHC GRS
1915+105 (see right panel of Fig 2) is probably the best-known
`microquasar' and is a strong source of hard X-rays, being well-detected by
BATSE over a six-year period. Foster et al. (1996) reported hard X-ray
states, or `plateaux' in this source. These plateaux seem to be
periods of steady jet production, like luminous variants on the
`canonical' Low/Hard X-ray state, and are generally followed by
discrete, relativistic ejection events (Fender et al. 1999a; Dhawan et
al. 2000). The GRANAT mission has also contributed immensely to our
understanding of patterns of hard X-ray emission from X-ray binaries
(e.g. Ballet et al. 1994; Churazov et al. 1994).  Many other examples
of hard X-ray detections of BHCs exist in the literature, but it seems
that the majority of detections are associated with the Low/Hard X-ray
state and/or transient outbursts, both of which are associated with
jet production (Figs 2 \& 3).

Grove et al. (1998) clearly establish that while the hard (photon
index $\sim 1.5$) X-ray power laws observed in the Low/Hard state tend
to break at about 100 keV, when the spectrum is dominated by a soft
(presumably disc) component the steep (photon index $\sim 2.5$)
power-law tail observed does not show a break to at least several
hundred keV. This weak, high energy tail in the soft state (when the
jet is apparently off - Fig 2) probably requires a nonthermal
population of Comptonising electrons (Poutanen 1998).

\section*{$\gamma$-rays : $>500$ keV}

Above $\sim 500$keV the number of detections of X-ray binaries drops
significantly. Cyg X-1, the classic BHC, was repeatedly detected at
$\geq 1$ MeV with COMPTEL on CGRO, and may even be stronger above 1
MeV in the `High/Soft' state (Fig 8; McConnell et al. 2001 a,b). The
relativistic jet source GRO J1655-40 and the Low/Hard state source GRO
J0422+32 are also both detected up to $\sim 1$ MeV (Grove et al. 1998).
More dramatically, Goldwurm et al. (1992) reported the detection of a
positron annihilation line from GRO J0422+32, although this may in
fact be associated with Lithium (e.g. Martin et al. 1994).

At even higher energies, Paredes et al. (2000 and references therein;
Fig 9) report the discovery of radio jets from the X-ray binary LS
5039, which is coincident with one of the unidentified EGRET
sources. They suggest that the $\gamma$-rays are produced by inverse
Compton upscattering of lower-energy photons by the nonthermal
relativistic electron population producing the resolved radio
emission. It is worth noting that LSI+61$^{\circ}$ 303 is also an
X-ray-faint, radio-bright system which may be associated with an EGRET
source (Harrison et al. 2000 and references therein).

\begin{figure}
\centering\epsfig{figure=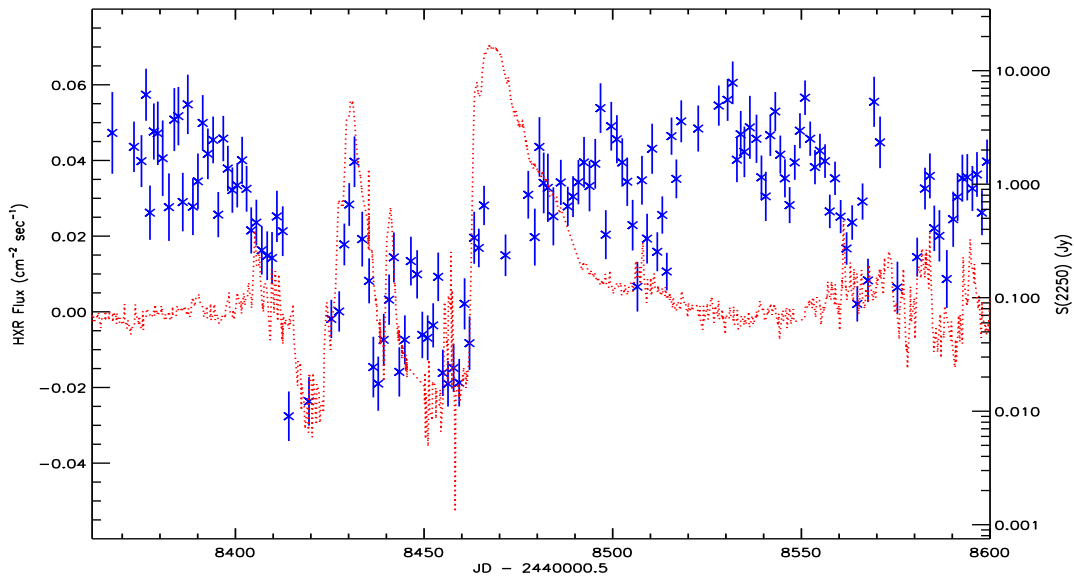,width=8cm,height=14cm,angle=90}
\caption{Bimodal hard X-ray [crosses] -- radio [dotted line]
(anti-)correlation in the jet source Cyg X-3. During `quiescent'
periods (e.g. JD 2448500-8550) hard X-rays and radio are
anti-correlated. During `active' periods, which constitute both flares
and pre-flare quenching, radio and hard X-ray fluxes are well
correlated (e.g. JD 2448420-8450). From McCollough et al. (1999).
}
\end{figure}

\section*{Discussion}

There seems to be a qualitative empirical relation between the
presence of hard (rather arbitrarily defined as 30--500 keV) X-ray
emission and radio emission in both NS and BHC X-ray binaries. Since
the evidence strongly points to the radio emission arising in a
synchrotron-emitting jet, this implies an association between the
presence of a hard X-ray emission and the production of an outflowing,
nothermal, population of relativistic electrons. Are these two things
related causally, or are they simply both more or less independent
manifestations of a particular accretion/outflow `state' ?

\begin{figure}
\centering\epsfig{figure=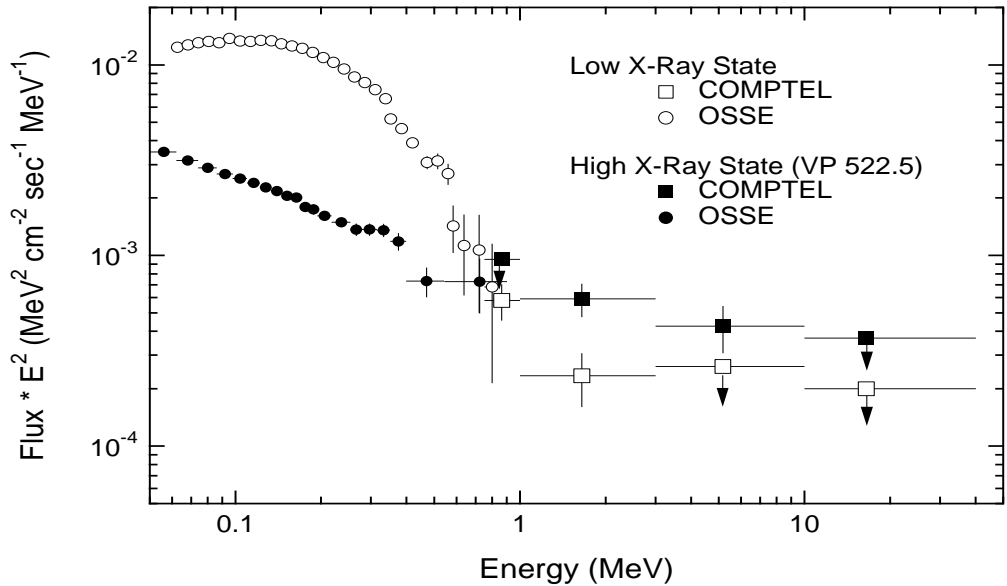,width=14cm,height=8cm,angle=0}
\caption{MeV spectrum of Cyg X-1 in the low/hard and high/soft X-ray states.
From McConnell et al. (2000a).
}
\end{figure}

Since (a) the hard X-ray emission is generally taken to arise in
(inverse) Comptonisation of soft photons by energetic electrons
(Sunyaev \& Titarchuk 1980; Poutanen 1998 and references therein), (b)
the synchrotron emission from the jet directly indicates a population
of energetic electrons, it might seem reasonable to guess that the
Comptonising electrons are part of the jet flow. In Fender et
al. (1999b) it was suggested that the `base' of the jet is responsible
for the Comptonisation in the canonical Low/Hard X-ray state. However,
the favoured models for the Low/Hard state invoke {\em thermal}
Comptonisation (Poutanen 1998 and references therein). 

Are the jets significant enough energetically to contribute in hard
X-rays, or is the power-law distribution responsible for the
synchrotron emission simply a small high-energy tail of a much more
energetically significant thermal distribuion ? Without accurately
normalised estimates of both populations for one source, it is hard to
make a comparison. However, Fender (2000a,b) argues that the jets from
BHCs in the Low/Hard state are likely to be a significant
power output channel for the accretion energy. If the synchrotron
spectrum extends merely to the near-infrared band ($\sim 2\mu$m) and
the jet has a radiative efficiency of $\leq 0.05$ then the jet is
likely to contribute $\geq 20$\% of the bolometric luminosity of the
system in this state. It seems therefore that in the Low/Hard state at
least there is enough power in the jet to justify consideration of its
contribution to the hard X-ray/$\gamma$-ray emission a source.

Aharonian \& Atoyan (1998) and Atoyan \& Aharonian (1999) discuss the
possibility of $\gamma$-rays from the jets of GRO J1655-40, GRS
1915+105 and SS 433. In particular, Atoyan \& Aharonian (1999) predict
that inverse Compton emission (or maybe even direct synchrotron
emission) directly from the jets in GRS 1915+105 could dominate the
high-energy emission above an MeV or so.
More radically, Markoff, Falcke \& Fender (2000) suggest that in the
Low/Hard X-ray state the broadband radio through X-ray spectrum is
dominated by the jet (with some contribution from thermal emission in
the UV and soft X-ray band). In this model the X-ray power-law is
dominated by optically thin synchrotron emission from the post-shock
region near the base of the jet. 

\begin{figure}
\centering\epsfig{figure=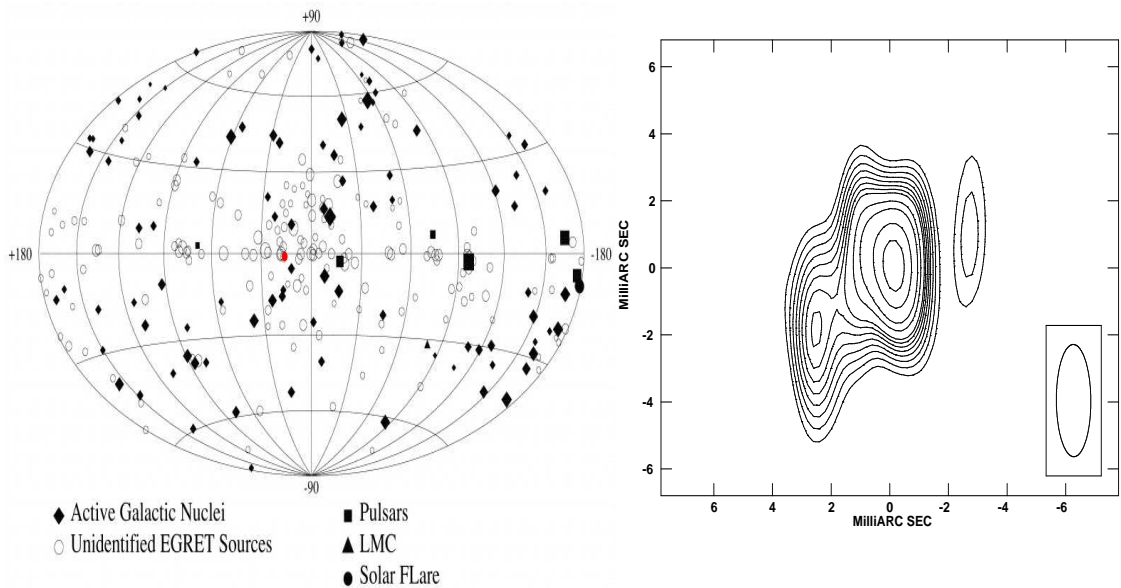,width=8cm,height=15cm, angle=270}
\caption{LS 5039 : a $\gamma$-ray binary ? The left panel shows the sky
distribution of unidentified EGRET $\gamma$-ray
sources. The weak X-ray binary LS
5039 is coincident with one of these (3EG J1824-1514). At radio wavelengths
LS 5039 is clearly resolved into an asymmetric (and therefore probably
relativistic) jet with the VLBA (right panel).
From Paredes et al. (2000).
}
\end{figure}

\section*{Conclusions}

To conclude, it seems that, while some famous and poorly-explained
exceptions exist in the literature (e.g. GRO J1655-40; Tavani et
al. 1996), in general the presence of `hard' X-ray emission (30--500
keV) is correlated with the presence of radio emission in both NS and
BHC X-ray binaries (with the exception of high-field X-ray pulsars
which do not seem to produce jets -- Fender \& Hendry 2000). In some
cases (e.g. GX 339-4; Fig 6) this correlation is very tight. Adopting
the classical picture of hard X-ray production via Comptonisation,
this broad correlation implies the simultaneous presence of a
Comptonising corona and synchrotron-emitting (and therefore
relativistic) jet in `hard' accretion `states'. This may suggest that
the `base' of the jet, within a few hundred Schwarzschild radii of the
compact object, is providing the hot electrons for the Comptonisation
process (Fender et al. 1999b), although models for Comptonisation in
the Low/Hard state of BHCs favour a thermal population of electrons in
the corona (Poutanen 1998 and references therein).  More radically,
the signature of the jet may extend to much higher energies than the
radio regime, and may even challenge the Comptonisation model of hard
X-ray production (Markoff et al. 2000).

The relation of jets to the presence of $\gamma$-ray ($>$ 500 keV)
emission is less clear, and further systematic studies (with
e.g. INTEGRAL) are required. However, given that radio jets, with
their population of very high energy electrons, are natural sites for
the production of high energy photons via both inverse Compton
scattering (Atoyan \& Aharonian 1999; Paredes et al. 2000) and maybe
even direct synchrotron emission (Markoff et al. 2000) such
connections should be investigated in detail in the future.

\section*{Acknowledgements}

I would like to thank Tiziana di Salvo, Didier Barret and Felix Aharonian for
reading through draft versions of this work.

\end{document}